\documentclass[english,a4paper,11pt]{scrartcl}
\usepackage[latin1]{inputenc}
\usepackage[dvips]{graphicx}
\usepackage{amssymb,amsmath}     
\usepackage{cite}
\makeglossary

\newcommand{\bc}{\begin{center}}
\newcommand{\ec}{\end{center}}
\newcommand{\bq}{\begin{quote}}
\newcommand{\eq}{\end{quote}}
\newcommand{\bi}{\begin{itemize}}
\newcommand{\ei}{\end{itemize}}
\newcommand{\bn}{\begin{enumerate}}
\newcommand{\en}{\end{enumerate}}
\newcommand{\bd}{\begin{description}}
\newcommand{\ed}{\end{description}}
\newcommand{\be}{\begin{eqnarray}}
\newcommand{\ee}{\end{eqnarray}}
\newcommand{\beq}{\begin{equation}}
\newcommand{\eeq}{\end{equation}}
\newcommand{\bs}{\begin{split}}
\newcommand{\es}{\end{split}}
\newcommand{\bdf}{\begin{definition}}
\newcommand{\edf}{\end{definition}}
\newcommand{\bt}{\begin{tabbing}}
\newcommand{\et}{\end{tabbing}}
\newcommand{\bfig}{\begin{figure}}
\newcommand{\efig}{\end{figure}}

\newcommand{\refsec}[1]{(Sec.~\ref{#1})}
\newcommand{\refeq}[1]{(\ref{#1})}
\newcommand{\reffig}[1]{(Fig.~\ref{#1})}
\newcommand{\reftable}[1]{(Table~\ref{#1})}

\newcommand{\LT}{LT$\alpha_1\beta_2$}

\newcommand{\MICRON}{$\mu\rm m$}

\begin{document}
\title{Modeling emergent tissue organization involving high-speed migrating cells in a flow equilibrium}
\date{\today}
\author{Tilo Beyer and Michael Meyer-Hermann\\
Frankfurt Institute for Advanced Studies,
Johann Wolfgang Goethe-University,\\
Max-von-Laue-Str. 1,
60438 Frankfurt Main, Germany\\[0.2cm]
Corresponding author: Tilo Beyer, E-mail: {\tt tbeyer@fias.uni-frankfurt.de}}

\maketitle
\begin{abstract}
There is increasing interest in the analysis of biological tissue,
its organization and its dynamics with the help of mathematical models. 
In the ideal case emergent properties
on the tissue scale can be derived from the cellular scale. However,
this has been achieved in rare examples only, in particular,
when involving high-speed migration of cells. One major difficulty
is the lack of a suitable multiscale simulation platform, which embeds
reaction-diffusion of soluble substances, fast cell migration and mechanics,
and, being of great importance in several tissue types, cell flow homeostasis. 
In this paper a step into this
direction is presented by developing an agent-based mathematical model specifically
designed to incorporate these features with special emphasis on
high speed cell migration. 
Cells are represented as elastic spheres migrating on a substrate
in lattice-free space. 
Their movement is regulated and guided by chemoattractants that
can be derived from the substrate.
The diffusion of chemoattractants is considered to be slower than
cell migration and, thus, to be far from equilibrium.
Tissue homeostasis is not achieved by the balance of growth and death
but by a flow equilibrium of cells migrating in and out of the tissue
under consideration. In this sense the number and the distribution of the cells
in the tissue is a result of the model and not part of the assumptions.
For purpose of demonstration of the model properties and functioning,
the model is applied to a prominent example of tissue in a cellular flow 
equilibrium, the secondary lymphoid tissue. The experimental data on cell speed
distributions in these tissues can be reproduced using reasonable mechanical
parameters for the simulated cell migration in dense tissue.
\\{\em keywords: agent-based model, multiscale model, chemotaxis, regular triangulation}
\end{abstract}

\maketitle
\section{Introduction}
There exists a number of agent-based off-lattice models to simulate tissue
\cite{Odell:1981,Palsson:2000,Palsson:2001,Dallon:2004,Drasdo:1995,Drasdo:2004,Honda:2004,Schaller:2005}.
Many of them describe tissue organization and pattern formation \cite{Meinhardt:1982}
by the dynamics of growth and death of cells. Other models are suitable for 
describing migration of a constant number of cells. 
The traditional research field of pattern formation involving migrating cells
\cite{Jiang:1998,Palsson:2001,Dallon:2004} is mostly based on a balance
of proliferation and cell death. 
Cells are migrating with relatively slow speed
with respect to other processes like diffusion of soluble substances.
Approximations based on slow cell migration are justified even for
such highly dynamic systems as the epidermis \cite{Meineke:2001,Schaller:2005}.

The model introduced
here is designed for tissue in a flow equilibrium of high-speed 
migrating cells as found frequently in immunological tissue.
Cells can enter and leave the tissue according to some dynamics imposed
by the considered biological system.
Feedback mechanisms between the cells and the substrate are incorporated
because homeostasis in such systems is often established by feedback
interaction \cite{Fu:1999,Ansel:2000,Nishikawa:2003}. 
This is modeled as transient excitation of a substrate or a small sessile cell population.
The excitation is thought to couple back to the cell migration thus allowing 
for a regulatory system in a stable flow equilibrium.

The model is based on a regular triangulation \cite{Edelsbrunner:1996,Okabe:2000}
that provides the cell neighborship topology which is changing rapidly due to fast
cell migration. The dual Voronoi 
tessellation \cite{Aurenhammer:1987,Okabe:2000} provides information about the shape of the cells
including cell contact surfaces \cite{Meineke:2001,Schaller:2005,Schaller:2006}.
The physical properties are based on previously introduced short-ranged elastic interactions
and actively generated intercellular forces
\cite{Drasdo:1995,Schaller:2005,Schaller:2006,Palsson:2001,Dallon:2004}. 
Each cell is modeled individually and can incorporate a set of molecular properties or
internal states that is appropriate for the system under consideration.
Such properties may have impact on cell migration or cell-cell and cell-substrate interaction.
The deterministic internal cell dynamics exhibit memory
in the sense that, for example, internal variables representing the cell
state like delay times are included.
Cell migration under the
influence of chemotaxis is also treated deterministic. 
However, cell orientation is stochastic
for unguided random cell migration.

In order to illustrate the model features, the model is used to investigate
certain features of cell behavior that are relevant for the formation of
primary lymphoid follicles (PLF). These can be found in secondary lymphoid tissue
(SLT)  like the spleen and lymph nodes
\cite{Fu:1999,Andrian:2003,Nishikawa:2003,Gunzer:2004,Cyster:2005,Finke:2005}.
Two lymphocyte cell populations of the PLF show fast migration:
B cells and T cells. Two sessile populations, namely follicular dendritic cells (FDC) and fibroblastic reticular cells (FRC)
serve as a complex 3D substrate for lymphocytes.
The morphology of SLT is characterized by PLF and T zones.
The PLF contains the B cells and the FDC. It is adjacent to the T zone harboring
T cells and FRC.
The functionality of the model features is demonstrated by analyzing
the migration of the lymphocytes and their interaction with chemokines.
Also the generation of a tissue in flow equilibrium
is studied using a toy model inspired by PLF formation.
On the basis of biologically proven interactions the model
can explain a clear separation of T and B cells
forming a T zone and an FDC containing follicle, as found in real tissue.
However, more realistic models of PLF remain to be developed in future work.

The presented simulation technique has a direct connection to experimental
constraints and exhibits the potential of building physically concise models of cell
migration in tissue. It's novelty with respect to previous work relies on the focus on
tissues composed of rapidly migrating cells which exhibit cell-flow-equilibrium.
This extends the range of applications of agent-based off-lattice techniques
to lymphoid tissue.
\section{Method}
\label{sec-method}
The tissue formation problem of migrating cells is simulated using an agent-based model
on top of a regular triangulation \cite{Edelsbrunner:1996,Okabe:2000,Schaller:2004,beyer:2005}.
The regular triangulation is used to provide the neighborhood topology for the cells that allows for
a continuous representation of cell positions and sizes in contrast to grid-based methods.
The regular triangulation also provides information about the cell contacts
and contact areas, and, 
within the limits of an approximation,
about cell volume and shape \cite{Aurenhammer:1991,Schaller:2005}. 
The regular triangulation will be briefly summarized in the next subsection.

The simulation of cells is realized in a 3-level multiscale model. 
The first level is the internal state of the
cells representing the dynamics of the phenotype of the cell
(see Sec.~\ref{sec-internal}). 
The second level models the contact interaction
between cells including mechanical interactions with the environment and exchange of signals by membrane
bound molecules (see Sec.~\ref{sec-motion}). 
The third level incorporates long range interactions via diffusive substances
for example chemotaxis, i.e.~directed motion of cells upwards a concentration
gradient which is induced by molecular chemoattractants
(see Sec.~\ref{sec-chemo}). 
Thus, chemoattractants are derived
from cellular sources and feed back to cell migration.
\subsection{Regular triangulation}
Each cell is represented as a sphere at position $\mathbf x$ with radius $R$. A vertex is defined
by the pair $X=(\mathbf x,R)$.
The regular triangulation is defined using the empty orthosphere criterion
\cite{Aurenhammer:1987,Edelsbrunner:1996,Okabe:2000}. In three dimensions four vertices $A,B,C,D$
forming a tetrahedron uniquely define an orthosphere \reffig{fig-regular}. The orthosphere
is empty if for any other vertex $V$
\be
 \left| \begin{array}{cccc}
       a_x - v_x & a_y - v_y & a_z - v_z & \left\|\mathbf a - \mathbf v\right\|^2  - R^2_a + R^2_v\\
       b_x - v_x & b_y - v_y & b_z - v_z & \left\|\mathbf b - \mathbf v\right\|^2  - R^2_b + R^2_v\\
       c_x - v_x & c_y - v_y & c_z - v_z & \left\|\mathbf c - \mathbf v\right\|^2  - R^2_c + R^2_v\\
       d_x - v_x & d_y - v_y & d_z - v_z & \left\|\mathbf d - \mathbf v\right\|^2  - R^2_d + R^2_v
 \end{array}\right| > 0\:
\ee
holds provided that the four vertices are oriented positively, i.e.~if
\be
  \left|\begin{array}{cccc}
    a_x & a_y & a_z & 1\\
    b_x & b_y & b_z & 1\\
    c_x & c_y & c_z & 1\\
    d_x & d_y & d_z & 1\\
\end{array}\right| > 0\:.
\ee
The symbols $a_x,a_y,a_z$ denote the coordinates of vertex $A$ at the position
$\mathbf a = (a_x,a_y,a_z)$ and $R_a$ the corresponding radius of the elastic sphere
associated with the vertex. The notation for the vertices $B,C,D,V$ is analogously defined.
A set of non-overlapping tetrahedras covering a set of vertices forms a regular 
triangulation if all 
orthospheres attributed to these tetrahedras do not contain any further vertex.
The regular triangulation, and with it the neighborhood topology,
changes when cells are moving, and when they are added or removed from the system
due to cell flux, cell death or cell proliferation. 
The corresponding algorithms have been developed and
published previously for serial \cite{Schaller:2004} and 
parallel computer architectures \cite{beyer:2005}.

\bfig
 \centering
  \includegraphics{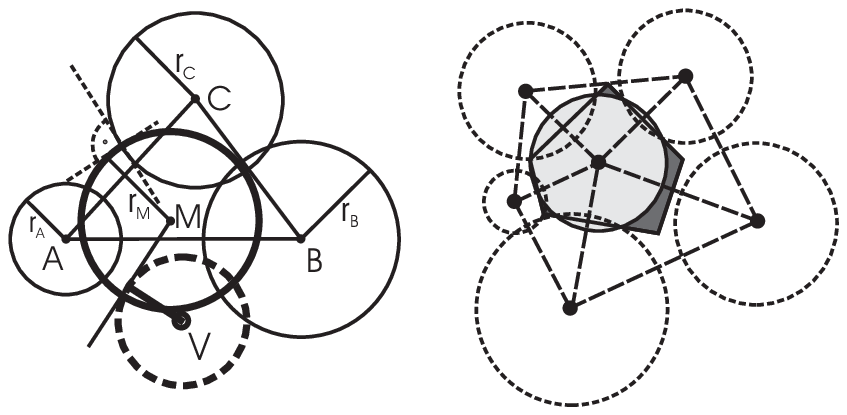}
 \caption{\label{fig-regular}
            Orthosphere in two dimensions (left panel). The triangle ABC defines an orthocircle M in two
            dimensions. The weights of the vertices are represented as disks with corresponding radii.
            Geometrically, the orthogonality of the orthocircle is indicated by the orthogonality
            of the tangents at the intersection point of the orthocircle and the circles representing the vertex
            weight. A vertex V is inside the orthocircle M when the tangential intersection lies within
            M, as shown in the example.
            A regular triangulation is marked as dashed straight lines connecting the vertices
            in the right panel. The dark grey polygon is the 
            Voronoi cell of the central vertex with an associated weighted sphere (grey disk). 
            The dashed
            circles depict the weighted spheres of the neighbor vertices. }
\efig
\subsection{Internal cell dynamics}
\label{sec-internal}
The phenotype of a cell is described by a set of internal cell variables $\phi$. These include internal times
to indicate when which type of event may happen. An example is the persistence time $T_{\rm p}$
of cells during chemotactic motion \cite{Miller:2002,Miller:2003,Wei:2003,Okada:2005,mehe:2005}.
When the persistence time $T_{\rm p}$ has past, the cell can reorient to the local chemoattractant field
\cite{Albrecht:1998,Ehrengruber:1996}.

Upon contact two cells can exchange signals via the contact surface. In a simple approach with unpolarized
ligand/receptor distributions the signal strength is proportional to the contact area. The contact area
is computed as the minimum between a sphere overlap and the common Voronoi face of these cells \reffig{fig-voronoi}.
This choice relies on the fact that the Voronoi-contact area is the better description for high density
of cells while the virtual overlap of the spheres is more realistic for low density systems
\cite{Schaller:2005,Schaller:2006}. In both cases the alternative measure for the shape leads to larger estimates
which justifies the present choice. In general the minimum of the two contact
areas is close to the realistic contact of two elastically deforming spherical cells.

\bfig
 \centering
  \includegraphics{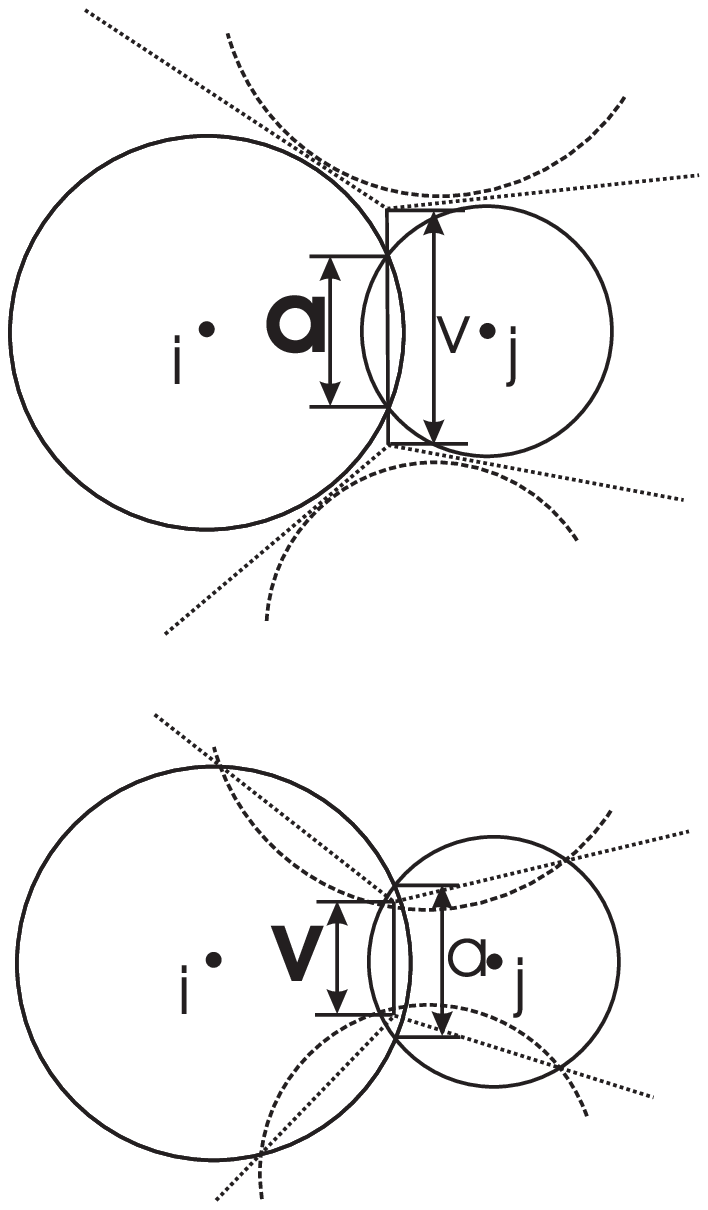}
 \caption{\label{fig-voronoi}
          Two-dimensional scheme of the contact surface of two cells. The upper panel shows the case
when the cells are loosely packed. The straight lines indicate the relevant part of the Voronoi tessellation.
The contact surface between the cells $i$ and $j$ is properly described by the overlap $a$ of the
two circles (the overlap of two spheres in three dimensions is a disk). The lower panel illustrates
the case of dense cell packing in which the Voronoi face $v$ is a better approximation for the
contact surface of the cells $i$ and $j$. Note, that this definition of the contact surface does
not only depend on the relative distance of the cells $i$ and $j$ but also on cells that
are common neighbors of these two cells (indicated by dashed circular lines).}
\efig

Finally, a set of variables describes the mechanics of the cell: velocity, orientation,
cell volume, and elasticity.
These variable couple directly to the next level of description, the contact interaction of cells, and have no
direct influence on other internal states but may influence the corresponding
variables of neighboring cells.
\subsection{Equations of motion}
\label{sec-motion}
The contact interaction of cells is predominantly given by mechanical 
interactions. It is described by
Newtonian equations of motion in the overdamped approximation
\cite{Palsson:2000,Dallon:2004,Schaller:2005}. In  this approximation acceleration of
cells and consequently conservation of moment can be ignored.
With dots denoting time derivatives we then have

\be\label{eq-newton}
 \bs
  0 &\approx m_i\mathbf {\ddot x}_i\\
    &= \mathbf F^{\rm act}_i\left(\phi_i\right)
      + \mathbf F^{\rm drag}_i\left(\mathbf {\dot x}_i,\{\mathbf {\dot x}_j\}_{\mathcal N_i}\right)\\
    & + \sum_{j \in \mathcal N_i}
           \mathbf F^{\rm act}_{ij}\left(\phi_i\right)
         + \sum_{j \in \mathcal N^{\rm c}_i}\left[ -  \mathbf F^{\rm act}_{ji}\left(\phi_j\right)
         + \mathbf F^{\rm pass}_{ij}\left(\mathbf x_i,\mathbf x_j\right)\right].
 \end{split}
\ee
The active forces $\mathbf F^{\rm act}$ 
on cell $i$ (if any) at position $\mathbf x_i$ depend on the internal state $\phi_i$ of 
the cell and the internal state of neighboring cells $\phi_j$,
while the passive forces $\mathbf F^{\rm pass}$ depend on the
positions of the cell $i$ and its neighbor cells $j \in \mathcal N^{\rm c}_i$
in contact with cell $i$. Active forces can also be exerted directly
to the surrounding medium. Therefore the set $\mathcal N_i$ includes
all Delaunay-neighbors $j$ independent of 
whether they are in physical contact with the cell $i$ or not.
All these forces are counter-balanced by the velocity-dependent drag 
forces $\mathbf F^{\rm drag}$. Because of the overdamped approximation
this results in an ODE system of first order for the cell positions.

The passive forces are composed of forces stemming from the cell's elasticity
and compressibility
\be\label{eq-all-passive}
  \mathbf F^{\rm pass}_{ij}\left(\mathbf x_i,\mathbf x_j\right)
=
\mathbf F^{\rm JKR}_{ij}\left(\mathbf x_i,\mathbf x_j\right)
+
\mathbf F^{\rm compress}_{ij}\left(\mathbf x_i,\mathbf x_j\right)
\quad,
\ee
where $\mathbf F^{\rm JKR}_{ij}$ recollects elastic interactions
and surface energy.

\subsubsection{Elastic cell-cell interaction}
Elastic forces between cells are treated according to the 
JKR-model \cite{Johnson:1971}. They depend on the virtual cell
overlap $h_{ij} = R_i + R_j - \left\|\mathbf x_i - \mathbf x_j\right\|$
where $R_i$ and $R_j$ are the cell radii.
\be\label{eq-JKR}
  \begin{split}
  F^{\rm JKR}_{ij}\left(\mathbf x_i,\mathbf x_j\right)
                     &= E_{ij}^* \sqrt{R_{ij}^*}\,h_{ij}^{3/2} - \sqrt{6\pi \sigma_{ij}E_{ij}^*{R_{ij}^*}^{3/2} h_{ij}^{3/2}}\\
  \frac{1}{E_{ij}^*} &= \frac{3}{4}\left[\frac{1-\nu_i^2}{E_i} + \frac{1-\nu_j^2}{E_j}\right]\\
  \frac{1}{R_{ij}^*} &= \frac{1}{R_i} + \frac{1}{R_j}
  \end{split}
\ee
with the cell's elastic moduli $E_i$ and $E_j$, Poisson numbers $\nu_i$ and $\nu_j$,
and the surface energy $\sigma_{ij}$. The force acts in direction of the normal
$\mathbf {\hat e}_{ij}$ on the contact face
\be
  \mathbf F^{\rm JKR}_{ij}\left(\mathbf x_i,\mathbf x_j\right)
    = F^{\rm JKR}_{ij}\left(\mathbf x_i,\mathbf x_j\right)\mathbf {\hat e}_{ij}
\ee
The applicability and the limits of the JKR-model for modeling elastic interactions
between spherical cells have been quantified recently \cite{Chu:2005}.

\subsubsection{Many-body interactions}
The JKR-model defines a two cell interaction which exhibits 
an equilibrium distance mediated by the balance of surface energy
and elastic repulsion. For two cells this leads to a negligible
deviation of the volume attributed to the vertex and the relaxed
volume of the real cell. However, cells will frequently interact
with several cells such that specific situations can occur 
in which cells are strongly compressed by surrounding
cells without a corresponding large relaxing force being generated.
Thus, cells might remain in a highly compressed state for too long
times because of the neglect of many-body interactions. 
To account for the cell volume and approximately
ensure volume conservation, a cell pressure concept is included \cite{Dallon:2004,Schaller:2005}. 
The pressure of cell $i$ is calculated as deviation of the actual cell volume $V_i$
from the target volume $V_i^*$
\be\label{eq-define-p}
  p_i &=& K_i\left(1 - \frac{V_i}{V_i^*}\right)\\
  K_i &=& \frac{E_i}{3(1-2\nu_i)}
\ee
where a linear compression model with compressibility $K_i$ is used. 
The forces resulting from this pressure are exerted between cells by adding the term
\be
  \mathbf F^{\rm compress}_{ij} = a_{ij}(p_i - p_j)\mathbf {\hat e}_{ij}   
\ee
to the passive cell forces $\mathbf F^{\rm pass}_{ij}\left(\mathbf x_i,\mathbf x_j\right)$
in \refeq{eq-all-passive}. 
$a_{ij}$ is the contact surface
of the cells with neighbor cells or the medium. In case of contact surface with the medium
$a_{ij}$ is calculated as minimum of the Voronoi face and the sphere overlap with an equal-sized
virtual cell being at equilibrium distance with respect to the JKR forces.
To get the actual volume the minimum of the sphere volume and the Voronoi cell volume is taken
which is motivated by similar arguments as for the contact area of two cells
\cite{Schaller:2005,Schaller:2006}.

\subsubsection{Force generation of migrating cells}
Previous models that incorporate the active forces of migrating cells \cite{Palsson:2000,Palsson:2001,Dallon:2004}
use a net propulsion force that is exerted into the direction of motion by pulling
on neighbor cells. However, cells also exert forces
perpendicular to the direction of motion which can be even larger than the net force
as it has been shown for keratocytes \cite{Burton:1999,Oliver:1999}. The target
application is lymphoid tissue which requires to describe lymphocytes. However, the mode of
migration is different for these cells. It depends on the so-called
constriction ring \cite{Lewis:1931,Murray:1992,Keller:1998,Wolf:2003,Paluch:2005}.
Therefore the active force generation model of fast migrating
cells is based on the constriction ring model 
thus taking into account the lateral forces during active migration of lymphocytes.

\begin{figure}
 \centering
  \includegraphics{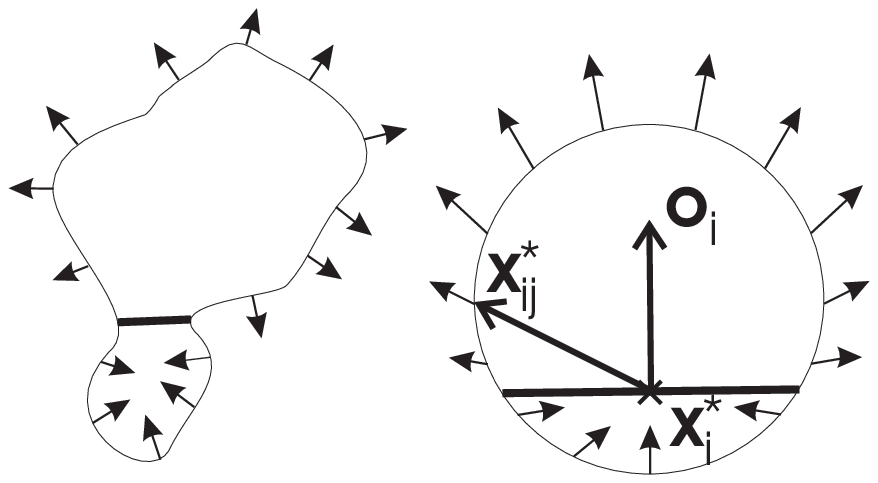}
 \caption{\label{fig-ring}
          Active forces of an actively migrating lymphocyte. A realistic force distribution
          of a migration lymphocyte is shown in the left panel with the constriction ring (thick line)
          separating the area from inwards directed forces from the area with outside directed forces.
          A mapping on a spherical cell representation is shown on the right.
          The force center is located in the center $\mathbf x^*_{i}$ of the constriction ring.
          The plane of the ring is perpendicular to the symmetry axis of the force generation determined by the
          cell's orientation $\mathbf o_i$.
          In the full tissue model the spherical surface is approximated by the
          polyhedral shape of a Voronoi cell determined by the neighbor cells $j$ positions such that the active
          forces acts at the contact points $\mathbf x^*_{ij}$.
          }
\end{figure}

Cells use a ring to attach a part of their membrane and cytoskeleton to the extra-cellular matrix and exert outward directed
pressure to their environment by contracting the rear of the cell \reffig{fig-ring}. The ring remains
fixed with respect to the extracellular matrix and therefor moves towards the end of a cell
during migration. In a simplified approach a
new ring is generated at the front of a cell only when the preceding ring has reached the end of the cell.
The force acting on cell $i$ by exerting active forces on a neighbor cell $j$ reads
\be\label{eq-ring}
  \mathbf F^{\rm act}_{ij} = a_{ij} p^*_i \,\mathrm{sign}[(\mathbf x^*_{ij} - \mathbf x^*_i)\cdot\mathbf o_i]
                             \,\frac{\mathbf x^*_{ij} - \mathbf x^*_i}{\|\mathbf x^*_{ij} - \mathbf x^*_i\|}
\ee
with the cell orientation $\mathbf o_{i}$, 
cell surface contact point $\mathbf x^*_{ij}$, constriction ring
center $\mathbf x^*_i$ (which generally differs from the cell's center),
interaction area $a_{ij}$ and the pressure $p^*_i$ actively exerted by cell $i$.
Note that $p^*_i$ is a parameter of the model and shall not
be confused with the pressure in \refeq{eq-define-p}.

Additionally a constant active force 
$-\mathbf F^{\rm act}_i\left(\phi_i\right)$ is directly exerted on the
extra-cellular matrix opposite to the direction of the cell's orientation
$\mathbf o_i$, i.e.~the cell is pushed forward
against the matrix by the force $\mathbf F^{\rm act}_i\left(\phi_i\right)$ in the direction $\mathbf o_i$.
The orientation $\mathbf o_i$ of the cell is parallel to the local chemotactic gradient
\refsec{sec-chemtax}.

The motion of cells can also be characterized using a motility coefficient that
is similar to diffusion coefficient in Brownian motion. However, it is not just a result
of physical contacts with the environment but also involves intrinsic features of cell migration.
The motility coefficient $D$ is influenced by the cell's speed $v$ and the mean free path length
\cite{Dickinson:1993}.
In cells the latter one is determined by the persistence time $T_{\rm p}$ during which
the cell maintains its polarized structured fixing the direction $\mathbf o_{i}$ of force exertion
and therefore the direction of motion \cite{Miller:2002,Miller:2003,Wei:2003,Okada:2005,mehe:2005}.
\be
  D=\frac{1}{3}v^2 T_{\rm p}
\ee
Thus the persistence time is a parameter describing the migration of cells
which is required in addition to the force generation.

\subsubsection{Friction forces}
The component of the friction forces between two cells $i$ and $j$ in contact read
\be
  \mathbf {\tilde F}^{\rm drag}_{ij} =
    \gamma_{ij} \left[\mathbf {\dot x}_{ij}  -  \mathbf {\hat e}_{ij}
    \left(\mathbf {\hat e}_{ij}\cdot\mathbf {\dot x}_{ij}\right) \right]
\ee
taking into account only the tangential parts of the 
relative velocity consistent with the treatment
of the cell-cell interaction by the energy conserving JKR-model \refeq{eq-JKR}. The force depends on the relative
cell velocity $\mathbf {\dot x}_{ij} = \mathbf {\dot x}_j - \mathbf {\dot x}_i$. 
The overdamped approximation and low Reynolds-numbers of cell migration justify the linear
velocity-dependence \cite{Palsson:2000,Dallon:2004,Schaller:2005}. 
The unit vector $\mathbf {\hat e}_{ij}$ is the normal vector of the plane of the cell contact.
The friction coefficient $\gamma_{ij}$ has the dimension of a viscosity times 
a length scale. 
$\gamma_{ij}$ is chosen proportional to the contact area between the cells. 
The total drag force is the sum
of all friction forces $\mathbf {\tilde F}^{\rm drag}_{ij}$ with neighbor cells plus a term of
the free surface of the cell having interaction with
the surrounding medium. The overall drag force is then given by
\be\label{eq-drag}
 \begin{aligned}
  \mathbf F^{\rm drag}_i &= -\gamma_{\rm med}\, \mathbf {\dot x}_i + \sum_{j\in \mathcal N_i} \mathbf {\tilde F}^{\rm drag}_{ij}\\
                         &=
    -\eta_{\rm med}R_i\left(1 - \frac{A_i}{A^{\rm tot}_i}\right) \mathbf {\dot x}_i\\
    &\quad+ \sum_{j\in \mathcal N_i} \left(\eta_i R_i+\eta_j R_j\right)\frac{a_{ij}}{A^{\rm tot}_i}
    \left[\mathbf {\dot x}_{ij}  -  \mathbf {\hat e}_{ij}
    \left(\mathbf {\hat e}_{ij}\cdot\mathbf {\dot x}_{ij}\right) \right]
 \end{aligned}
\ee
with medium viscosity $\eta_{\rm med}$ and the cell-specific viscosities $\eta_{i}$.
$A_i = \sum_{j\in\mathcal N^{\rm c}_i} a_{ij}$ is the surface in contact with other cells.
$A^{\rm tot}_i=\sum_{j\in\mathcal N_i} a_{ij}$ is the total surface of a cell.
The friction coefficient fulfills the symmetry $\gamma_{ij} = \gamma_{ji}$.
The form of the friction coefficients is motivated by the Stokes relation for the friction
of a sphere at velocity $\mathbf v$ in a medium with viscosity $\eta$
\be
  \mathbf F^{\rm Stokes} = 6\pi\eta R \mathbf v.
\ee
The geometry related factor $6\pi$ is absorbed in the cell-specific viscosity $\eta_i$ in \refeq{eq-drag}.
The coefficients $\eta_{\rm med}$ and $\eta_i$
are chosen such that reasonable active forces have to be generated by the cells in order
to match the cell velocities measured experimentally (see Table \ref{tab-par}).
\subsection{Reaction-diffusion system of chemoattractants for long-range interactions}\label{sec-chemtax}
\label{sec-chemo}
The chemotaxis of cells is described by coupling the direction of migration
$\mathbf o_i$ of a cell $i$ to the
local chemoattractant gradient. According to the observation that 
leukocytes tend to have a persistence time $T_{\rm p}$
between subsequent orientation changes 
\cite{Wei:2003,Gunzer:2004,Okada:2005,Miller:2002,Miller:2003}
the gradient is sensed by the simulated cells periodically and 
$\mathbf o_i$ is kept constant in between.
The concentration of a chemoattractant is computed solving the time dependent diffusion equation.
The time dependence is included for two reasons. First, to account for a dynamics of sources that frequently
generates new or removes present sources. The
time to equilibrate the concentration after changing the source is far longer than the typical
times for cell migration such that the cells can sense this temporal concentration change.
Second, the internalization dynamics presented in the next section acts on a time scale comparable to the cell
migration but much faster than the time scale required for the chemokine diffusion
to be equilibrated on the tissue scale.

\subsubsection{Internalization model}\label{sec-int}

The internalization of chemoattractant receptors occurs naturally when they are bound by their ligand
\cite{Arai:1997,Moser:2004,Neel:2005}.
Studies showed that desensitization of chemoattractant receptors 
exposed to high chemoattractant concentrations \cite{Tomhave:1994} occur probably by
internalization. Alternative possibilities would be a chemical modification of the receptors which
is not considered here.
Of note, some experiments fail to detect chemoattractant responses
of freshly isolated cells despite the presence of the corresponding receptor \cite{Roy:2002}.
Thus, the presence of receptors is necessary but not sufficient to cause chemotaxis of cells.
Most likely the suppression of the receptor function is mediated by a desensitization
mechanism either by internalization or cross-desensitization \cite{Moser:2004}.

Desensitization of cells with respect to a chemoattractant $c$
is achieved by internalization of the receptors 
which have bound the chemoattractant
with rate $k_{\rm i}$ \cite{Neel:2005}.
Thus the receptor comes in three states: free on the cell membrane ($R$), 
bound to the ligand on the cell membrane ($R_b$), and
internalized with the ligand ($R^*$) \reffig{fig-desense},
leading to the dynamics:
\be\label{eq-desens}
 \bs
  \dot R         \quad&=\quad - k_{\rm on}R\,c + k_{\rm off} R_{\rm b} + k_{\rm r} R^*\\
  \dot R_{\rm b} \quad&=\quad k_{\rm on}R\,c - k_{\rm off} R_{\rm b} - k_{\rm i} R_{\rm b}\\
  \dot R^*       \quad&=\quad k_{\rm i} R_{\rm b} - k_{\rm r}R^*\\
  \dot c         \quad&=\quad - k_{\rm on}R\,c + k_{\rm off} R_{\rm b}
  - \kappa c + Q + \Delta c
 \end{split}
\ee
The binding of the ligand is characterized by the on and off rate 
constants $k_{\rm on}$ and $k_{\rm off}$ leading to
the ligand-receptor complex $R_{\rm b}$. The rate
$k_{\rm r}$ describes the recycling of the internalized receptor $R^*$ 
into the free membrane form $R$. The basic
assumption is that the total receptor content $R_{\rm tot} = R + R_{\rm b} + R^*$ 
is conserved, i.e.~no terms
describing the transcription or degradation of the receptor 
are included in \refeq{eq-desens}. This view is supported
by the fact that internalized ligand-receptor complexes dissociate such
that free receptor become available in the cell \cite{Neel:2005}.
The assumption allows to eliminate the equation for $R_{\rm b}$ from the
system \refeq{eq-desens}.

\bfig
 \centering
  \includegraphics{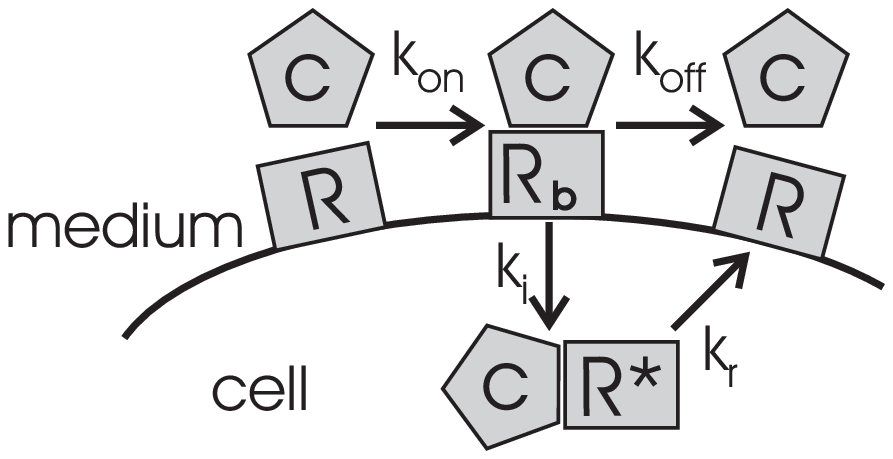}
 \caption[Receptor desensitization model]{\label{fig-desense}
          Receptor desensitization model. The chemoattractant $c$ binds with the rate constant $k_{\rm on}$
          to its receptor $R$. The receptor-chemoattractant complex can either dissociate with the rate
          constant $k_{\rm off}$ or
          internalized with the rate constant $k_{\rm i}$. The internalized complex gets recycled re-expressing
          the internalized receptor $R^*$ on the surface. The recycling is characterized
          by the rate constant $k_{\rm r}$ }
\efig

The reaction equations \refeq{eq-desens} are completed by an unspecific decay $\kappa c$
of the chemoattractant, a source term $Q$, and a diffusion term $\Delta c$ for the ligand.
The decay term is in accordance with the fact that chemoattractants are inactivated 
and processed by all cells without
reaction with the proper receptor \cite{Moser:2004}.
This limits the life time of a chemoattractant molecule. 
The receptors are also transported by cells 
such that the whole system \refeq{eq-desens} is a 
reaction-diffusion system. It is solved using a splitting method 
of first order in time \cite{Karlsen:1999,Tyson:1996}.

\section{Model of primary lymphoid follicle formation}
This section briefly describes the underlying biological concepts of PLF dynamics.
The required ingredients to study the formation of the PLF are the B and T cell flow
and the dynamics of the sessile cell populations,
i.e~the behavior of FDC and FRC. The parameters
for the simulation are given in table \ref{tab-par}. A detailed parameter estimate for the
reaction-diffusion system \refeq{eq-desens} is given in the appendix.

\bfig
 \centering
  \includegraphics{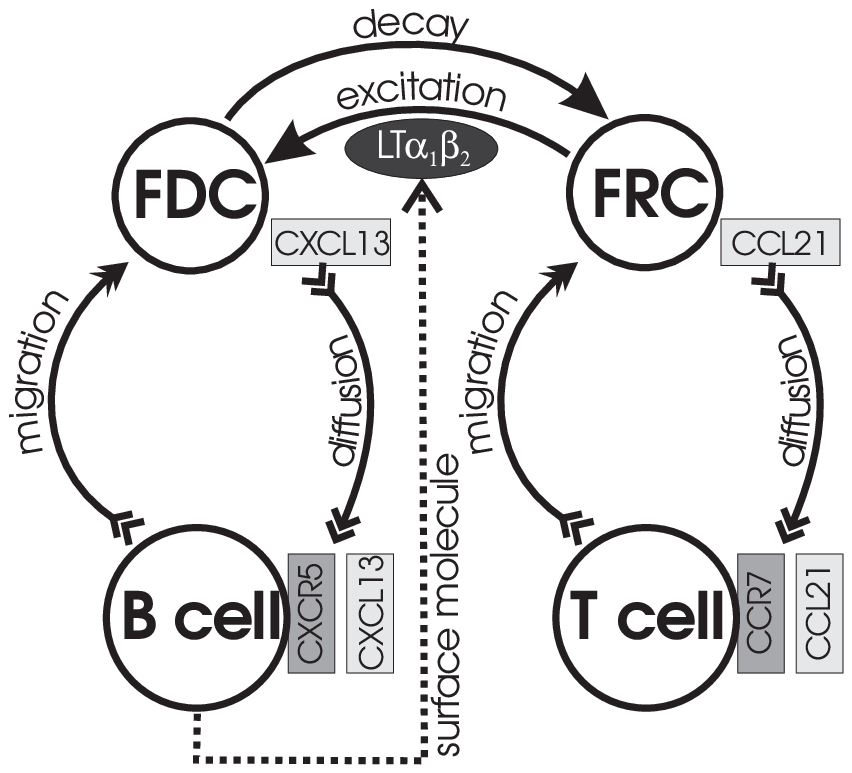}
 \caption{\label{fig-mig}
          Interaction network regulating migration in the PLF system. The chemoattractants
          CXCL13 and CCL21 guide the migration of B and T cells via the receptors CXCR5 and CCR7,
          respectively. The chemoattractants are produced by different non-migrating cells.
          Thereby FRC can become FDC when the are excited by a B-cell derived signal (\LT). FDC return
          to the FRC state when the signal is absent.}
\efig

\subsection{Flux of B and T cells}\label{sec-flux}
B cells are constantly entering SLT predominantly via specialized blood vessels
(reviewed in \cite{Cyster:2005}).
It is less clear how B and T cells leave this tissue. Recent experimental
data suggests that lymphatic vessels guide cells to leave SLT \cite{Cyster:2005}.
In a simplified model cells enter and leave the tissue through separate
spherical areas that do not further interact with the system.
However, one should be aware that in real SLT the lymphatic endothelium which
may be involved in the efflux of cells is located exclusively outside the PLF
\cite{Belz:1998,Ewijk:1980,Pellas:1990,Belisle:1990,Azzali:2000,Azzali:2002,Azzali:2003,Drinker:1933,Ohtani:1986,Ohtani:1991,Belz:1995}.
Thus, there may be some interaction that couples PLF formation with
the location of exit spots which we ignore in the presented toy model.

In order to allow the cells to remain and interact within the tissue for a certain time
they are not permitted to use
the exit areas before a minimal transit time has past. This time is about 3 hours according
to a receptor dynamics \cite{Lo:2005}. This is consistent with measurements of the minimal
transit time of lymphocytes \cite{Pellas:1990,Pellas:1990b,Young:1999}. In order to establish
a sufficient cell efflux a chemoattraction towards the exit spots is assumed. This is
supported by experimental evidence \cite{Lo:2005,Schwab:2005,Cyster:2005}.
Due to a lack of information the corresponding receptor is treated with a simplified model
without taking into account receptor internalization.

\subsection{Migration of B and T cells}
The migration of B and T cells within SLT follows similar principles. 
Both cells can perform chemotaxis
and random motion. It has been shown experimentally that
B cells are predominantly attracted by the chemoattractant CXCL13 \cite{Cyster:2005}.
Both, B cells and T cells respond to CCL21 with T cells being more sensitive \cite{Cyster:2005}.
The corresponding receptors are CXCR5 (for CXCL13) and CCR7 (for CCL21).
FRC are the source for CCL21 and FDC produce CXCL13 (FIG.~\ref{fig-mig}).

\subsection{Dynamics of sessile cell populations}
Although the origin of FDC is not clear there is strong evidence that they
are derived from FRC found in the T zone of SLT \cite{Nierop:2002}.
Therefore, it is assumed that FDC
are an excited state of FRC. The excitation is mediated by the contact signal \LT~provided
by the B cells \cite{Fu:1999,Ansel:2000,Nishikawa:2003} (FIG.~\ref{fig-mig}).
The FDC-B cell interaction has been shown to be reversible \cite{Fu:1999} in the sense
that the absence of the B cell signal lets the FDC vanish. In the model context that is described
as the decay of the excited FDC state back into the FRC state.
Note that, alternatively to the assumed scenario FRC and FDC may also have a common
progenitor which can develop in either FRC or FDC depending on external stimuli. 
This is not considered explicitely in the present model.

Within the PLF model an FRC differentiates to an FDC when the signal 
threshold for \LT~has been exceeded for
a given time $T_{\rm FRC\leftarrow FDC}$. 
The signal is determined by summing up all \LT~contributions
from neighbor cells, i.e.~surface density of \LT~times contact area. 
The differentiation is then instantly
performed changing the internal cell states of an FRC into that of an FDC, 
i.e.~replacing a source
for CCL21 by a source for CXCL13. In a similar manner FDC
differentiate back to FRC after the \LT~signal is below the 
threshold for a critical time $T_{\rm FDC \leftarrow FRC}$.
For simplicity and 
considering the lack of corresponding experimental data 
$T_{\rm FRC \leftarrow FDC} = T_{\rm FDC \leftarrow FRC}$
and equal \LT~thresholds are assumed.

\subsection{Sequence of events for follicle formation}
The overall picture of PLF formation is as follows. A background of immobile FRC produces CCL21.
B and T cells enter the tissue through the blood vessels which are represented by few ($<$10)
spheres of 30 \MICRON~diameter \cite{Grayson:2003} randomly scattered in an area of 150 \MICRON~size.
When sufficiently large B cell aggregates
form, FRC are induced to become FDC by \LT~thus replacing 
a production site for CCL21 by a CXCL13 source.
This attracts more B cells and enlarges
the forming PLF.
T cells are kept outside the PLF just by
coupling to the CCL21 which is produced around the PLF by the remaining FRC.
Both, B and T cells, leave the tissue through 
exit spots which are represented as spheres
acting as sinks for cells.
Any dynamics on the exit spots that may result from PLF formation
are ignored (compare Sec.~\ref{sec-flux}).

\begin{table}
 \centering
 {\small
 \begin{tabular}{lrll} \hline
   {\bf parameter} & {\bf value} & & {\bf remarks/ref.} \\ \hline
   B/T cell diameter                   & 9 \MICRON~    & & \cite{Ewijk:1980,Thompson:1984,Bornens:1989,Haston:1984} \\
   $E_i$               & 1 kPa & & \cite{Bausch:1998,Bausch:1999,Forgacs:1998,Chu:2005} \\
   $\nu_i$               & 0.4   & & \cite{Maniotis:1997,Hategan:2003} \\
   $\sigma_{ij}$          & 0--0.3 ${\rm nN}\,\mu{\rm m}^{-1}$ & & \cite{Moy:1999,Verdier:2003} \\
   $T_{\rm p}$                  & 120--180 s & & \cite{Miller:2002,Wei:2003} \\
   $\eta_i,\,\eta_{\rm med}$             & 500 nN$\,\mu{\rm m}^{-2}\,s$ & & \cite{Miller:2002,Miller:2003,Wei:2003,Okada:2005,Bausch:1998,Bausch:1999,Lauffenburger:1996}\\
    
   \hline
   $T_{\rm FRC \rightarrow FDC}$ & 3 h & & \cite{Ewijk:1980,Mackay:1998,Mebius:1997} \\ 
   
   \hline
   
   $D$                              & 10--100 $\mu{\rm m}^2\,s^{-1}$ & & \cite{Randolph:2005} \\
   size of diffusion grid                          & 1200 \MICRON~& & \\
   grid resolution                    & 35 \MICRON~ & & \\
   max.~cell displacement $\Delta x$ & 0.9 \MICRON~  & &  \\
   min.~time resolution $\Delta t$   & 10 s         & & \\
   B  : T cell ratio                & 0.4 : 0.6    & & \cite{Young:1999,Sacca:1998}\\
   influx of cells (B + T)              & 0.1--2 ${\rm cells}\,s^{-1}$ & & \cite{Young:1999,Sacca:1998} \\
   size of simulation area          & 600 \MICRON~ & & \\
   number of FRC                        & 2500         & & \\
   
   \hline
 \end{tabular}
 }
 \vspace{0.3cm}
 \caption[Common model parameters]{\label{tab-par} The parameters used in the simulation for the PLF.
          The references given support the used value. 
          If no references or comments are given, the parameter is a systemic model 
          parameter. The value is chosen to guarantee sufficient accuracy of the simulation.
          The parameters for the reaction part of the system \refeq{eq-desens} are estimated
          in detail in the appendix.}
\end{table}
\section{Results}
\subsection{Biomechanics of lymphoid tissue}

\bfig
 \centering
 \includegraphics[angle=-90,width=8cm]{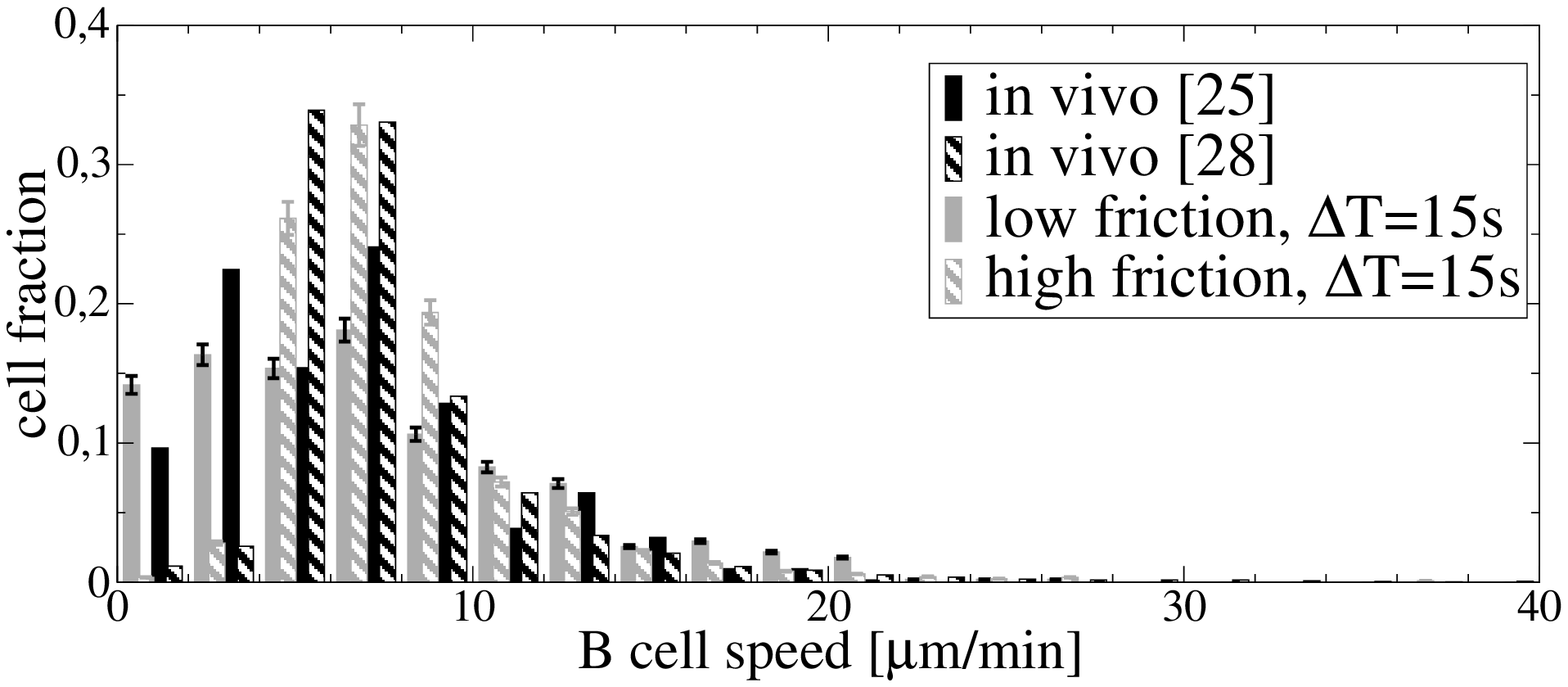}
 \includegraphics[angle=-90,width=8cm]{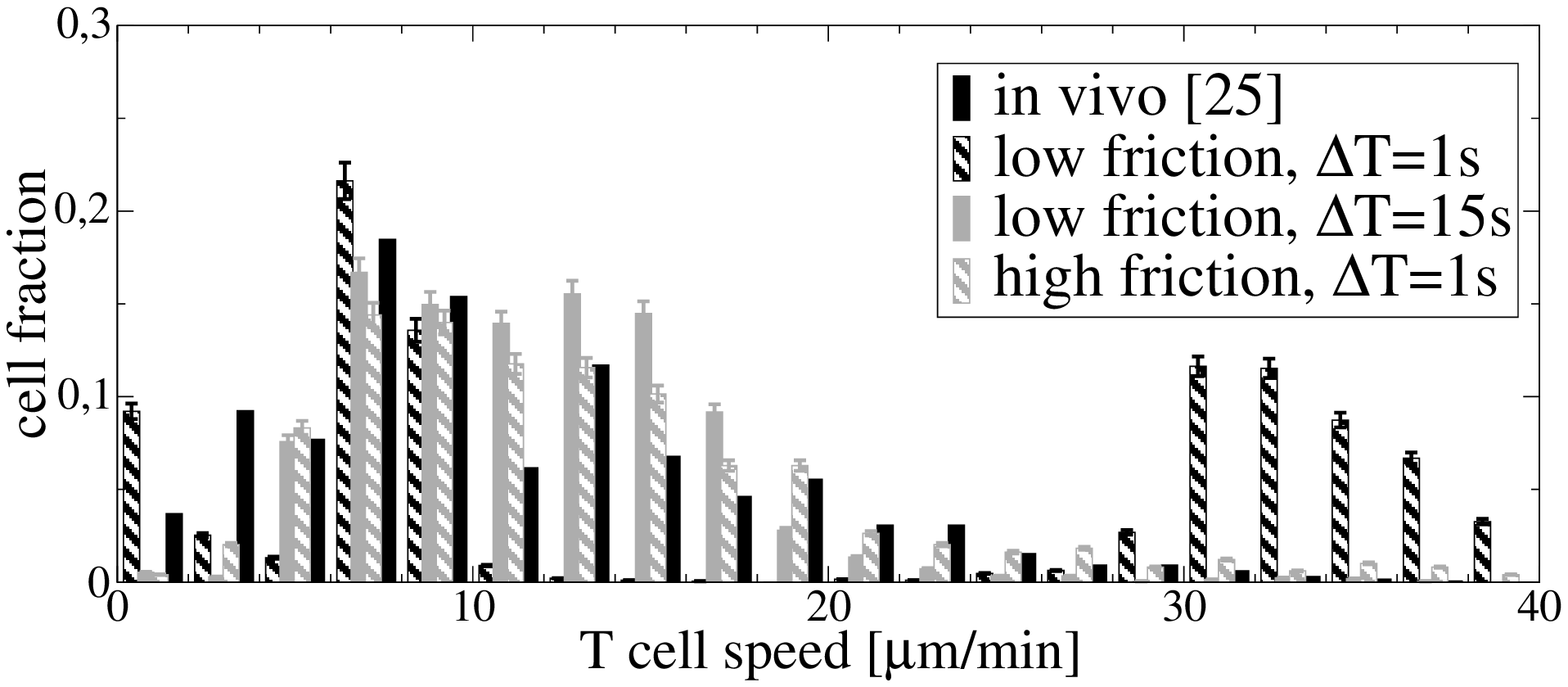}
 \includegraphics[angle=-90,width=8cm]{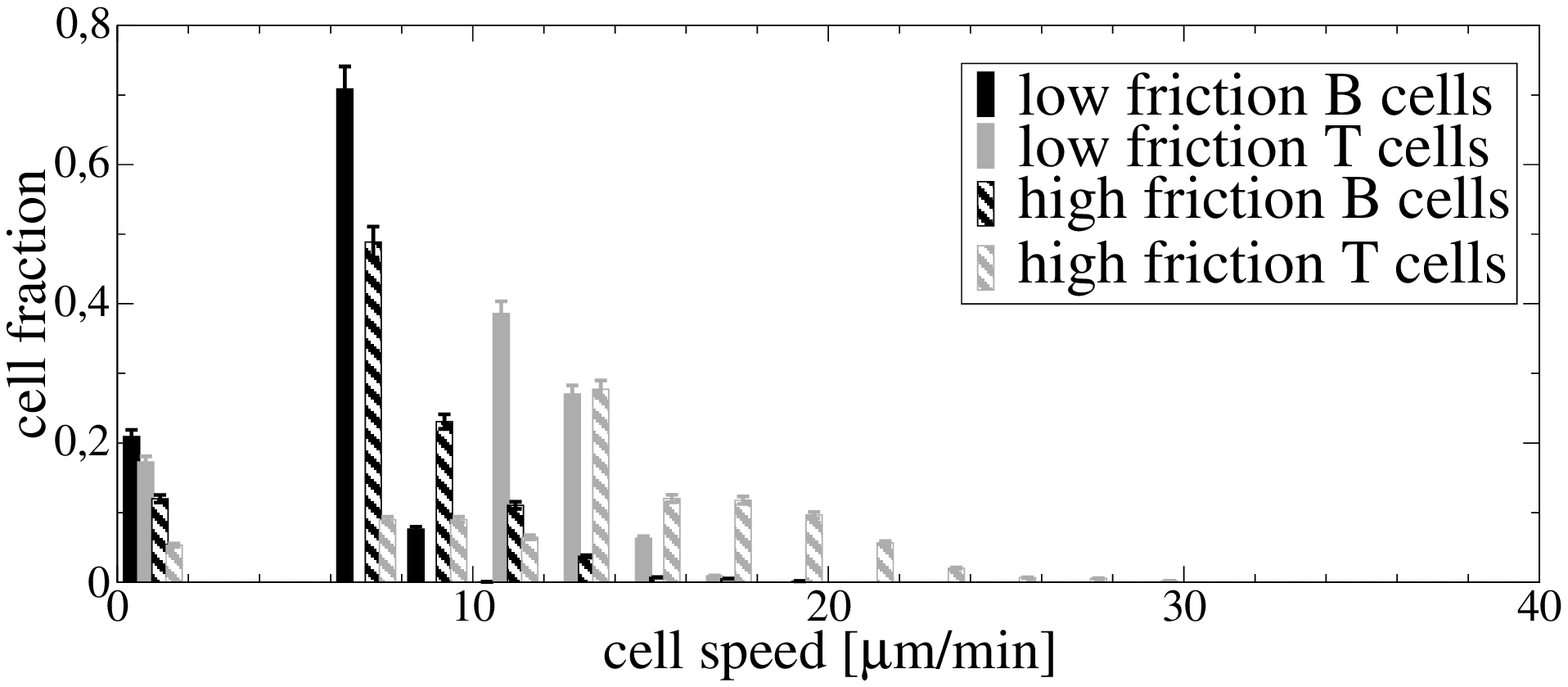}
 \caption{\label{fig-biomech}
    Comparison of experimental and simulated cell speed distributions. Each bin represents an
    2 $\mu{\rm m\,min^{-1}}$  speed interval. For clear visibility the bins are placed
    around the centers of the speed intervals. The top left panel shows the B cell speed distribution
    at low ($F^{\rm act}_{\rm B cell}=18\pm3\,{\rm nN}$, $p^*_{\rm B cell}=0.04\pm0.01\,{\rm nN\,\mu{\rm m}^{-2}}$)
    and high friction ($F^{\rm act}_{\rm B cell}=120\pm20\,{\rm nN}$, $p^*_{\rm B cell}=1.7\pm0.05\,{\rm nN\,\mu{\rm m}^{-2}}$)
    compared to two experimental data sets. The top right panel
    shows the speed distribution of T cells. The way of determining the speed distribution is crucial in the case of the low
    friction ($F^{\rm act}_{\rm T cell}=22\pm3\,{\rm nN}$, $p^*_{\rm T cell}=0.06\pm0.02\,{\rm nN\,\mu{\rm m}^{-2}}$). With a high
    sampling rate ($\Delta T=1\,\rm s$) the distribution has a contribution at high speeds ($>30\,\mu{\rm m\,min^{-1}}$) that is not
    seen using a larger sampling interval ($\Delta T=15\,\rm s$). For comparison the high friction regime
    ($F^{\rm act}_{\rm T cell}=240\pm35\,{\rm nN}$, $p^*_{\rm T cell}=3.7\pm1.1\,{\rm nN\,\mu{\rm m}^{-2}}$)
    is also shown (in that case the speed distributions with $\Delta T=1\,\rm s$ and $\Delta T=15\,\rm s$ are virtually
    indistinguishable).
    When the inter-cellular active forces are omitted ($p^*_{\rm T cell}=p^*_{\rm B cell}=0$ \refeq{eq-ring}) the speed distributions
    show rather sharp peaks (lower panel, sampling interval $\Delta T=15\,\rm s$).
 }
\efig

The active forces exerted by lymphocytes to the ECM ($F^{\rm act}_i\left(\phi_i\right)$) and to other cells
($p^*_i$, \refeq{eq-ring}) are not known explicitely (there are just estimates based on experiments of other cell types
\cite{Balaban:2001,Burton:1999,Galbraith:1999,Schwarz:2002b}). The same applies to the friction parameters
($\eta_i$ and $\eta_{\rm med}$, \refeq{eq-drag}). 
For simplicity it is assumed that 
$\eta_i=\eta_{\rm med}$ for all lymphocytes.
As the known average speed is essentially
determined by the ratio of active forces and the friction parameters
only 2 parameters per cell type remain unknown. 
These can be determined from the measured speed distributions
for B and T cells within SLT \cite{Miller:2002,Okada:2005}. 
A reduced cell system with a constant number of lymphocytes
that respond to a chemotactic point source is used.
The cell constituents are either B or T cells only or an even mix of both cell types.

The parameters for the JKR-forces are relatively well known.
The parameters entering \refeq{eq-JKR} are the elasticity of lymphocytes $E_i=1 kPa$  
\cite{Bausch:1998,Bausch:1999,Forgacs:1998,Chu:2005},
the Poisson number $\nu_i=0.4 $ \cite{Maniotis:1997,Hategan:2003},
and the surface energy $\sigma_{ij}=0--0.3 {\rm nN}\,\mu{\rm m}^{-1}$
\cite{Moy:1999,Verdier:2003}.
Thus the elastic cell interactions can serve as reference
forces limiting the reasonable range of friction parameters.
Therefore the fit to the speed distribution was performed 
with two qualitatively different friction regimes.
In the low friction regime  
($\eta_i=\eta_{\rm med}=50\,{\rm nN}\,\mu{\rm m}^{-2}\,s$) the active forces that are required
to produce the correct speed distributions are of the same
order as the JKR forces.
In this regime it turns out that the mode of measuring
the cell speed becomes crucial:
The speed distribution of the real system was measured
on the basis of the average displacement over an interval
of $\Delta T=10-15\,\rm s$ \cite{Miller:2002,Okada:2005}.
These data are reproduced and compared to the simulation
results in the two upper panels of FIG.~\ref{fig-biomech}.
Using a faster sampling with $\Delta T=1\,\rm s$ the speed distribution
becomes different and exhibits the influence
of JKR forces that lead to fluctuations at rather high speed (FIG.~\ref{fig-biomech}
top right panel).
The JKR forces dominantly lead to reversed or perpendicular motion of cells with respect
to their migration axis when cells collide due to active movement.
Thus the net-displacement by JKR forces during $\Delta T=15\,\rm s$ is rather small and does 
only show up at lower speeds in the
resulting distribution when longer sampling intervals are used.

In the second regime with high friction
($\eta_i=\eta_{\rm med}=500\,{\rm nN}\,\mu{\rm m}^{-2}\,s$) (FIG.~\ref{fig-biomech})
the average speed with $\Delta T=1\,\rm s$ and $\Delta T=15\,\rm s$ match each other
(not shown). The resulting speed distribution
usually fit better to the experimental ones, in particular, because 
low speed contributions now are shifted to
the peak resulting in a sharper speed distribution.
Moreover the components of the 
speed distribution in the low speed range are smaller 
as the active forces can more easily overcome the
JKR forces such that cells are less likely to be immobilized 
in an cellular environment of high density.
However, the active forces are higher than measured experimentally
\cite{Balaban:2001,Burton:1999,Schwarz:2002b}.
Yet, the parameter set is computationally favorable 
as the high speed fluctuations are missing. This allows for
larger time steps in the simulation.

The effect of active forces $F^{\rm act}_{ij}$ resulting from the constriction ring model
is relevant to the speed distribution.
Without these active inter-cellular forces the speed distribution shows only sharp peaks (FIG.~\ref{fig-biomech} lower panel).
Also when the speed distribution of each lymphocyte subset is fitted using a homogeneous
cell aggregate the peaks in the speed distribution tend to be sharper than in
experiment. This could be compensated
with larger active cell interaction forces $F^{\rm act}_{ij}$ which however produces the
wrong speed distribution when B and T cells interact which each other (not shown).

\subsection{Stable follicle formation}
\begin{figure*}
 \centering
  \includegraphics{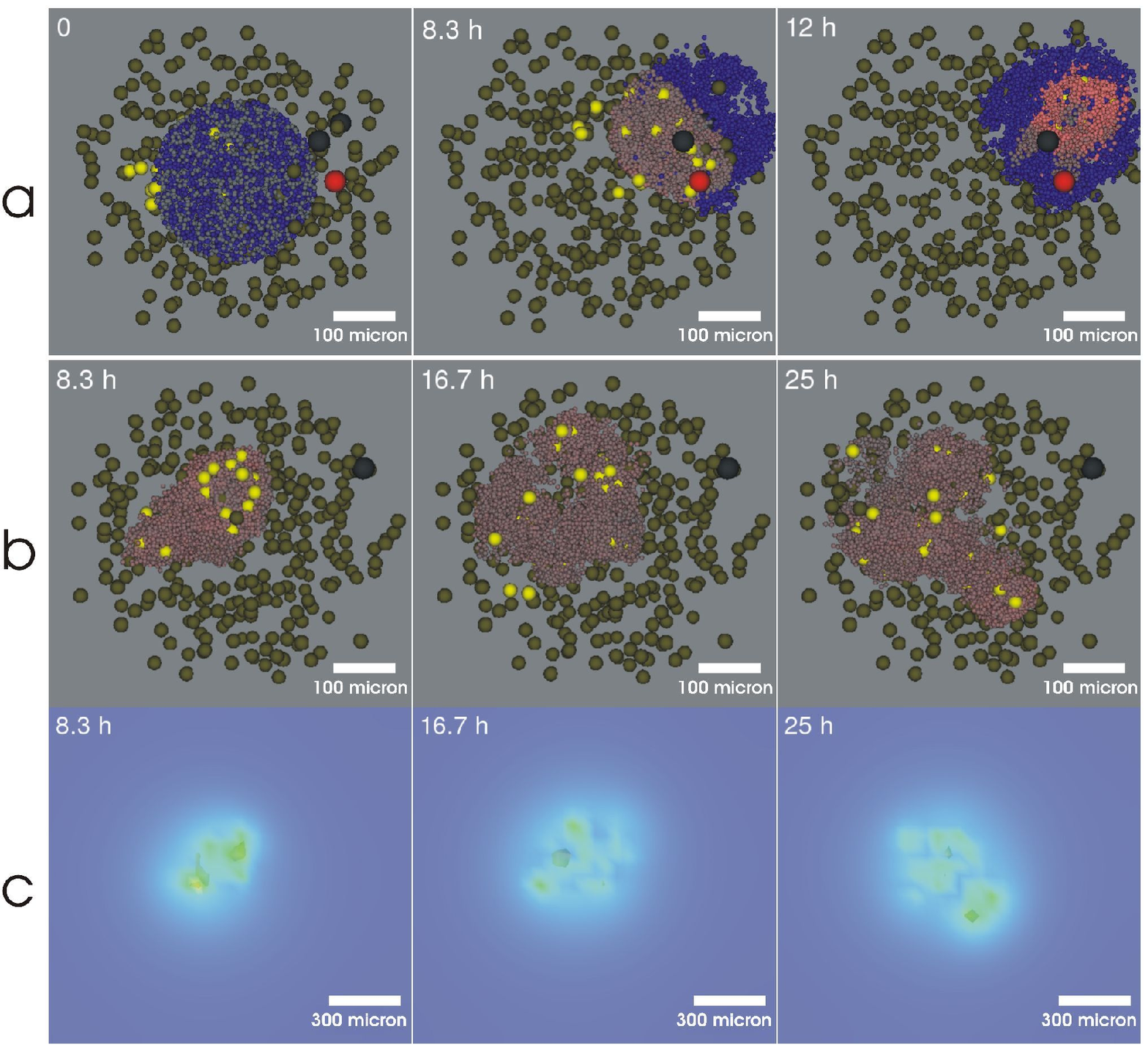}
 \caption{\label{fig-morph}
          (Color online) Three-dimensional slice projection of the simulated tissue (a,b).
          The images have been rendered with POVRay (http://www.povray.org).
          FRC are shown as medium sized dark grey spheres (olive),
          FDC as bright medium size spheres (yellow), and T cells as small dark spheres (blue).
          B cells are shown as small grey spheres (grey to red) (a,b).
          Cells can enter the simulated tissue through blood vessels shown as large grey (red) spheres and exit through
          exit spots indicated as large black (black) spheres. Colors given in brackets refer to the
          colored online version of this figure. \newline
          The localization of the follicle is stable (a). Even a preformed FDC network with a number of
          unsorted B and T cells
          cannot prevent the formation of the follicle around the exit spots. The remainder
            of the initial configuration is an orientation of the PLF in the direction of the initial FDC
            (a, 8.3 h) which is dissolved later on (a, 12 h). \newline
            The internalization dynamics is visualized in a simulation
            with a constant number of B cells (b).
            Receptor internalization is found to be a source of instability
            involving rapid morphological changes of B cell distributions.
            Note that the CXCL13 distribution behaves accordingly 
            (c, CXCL13 concentration increases from dark grey (blue) to white (red)).
            The concentration Was visualized with OpenDX (http://www.opendx.org).
 }
\end{figure*}

The simple model of PLF formation is able to generate stable follicle sizes of a few 100 \MICRON~diameter
with roughly $10^4$ B cells. The follicle forms around the exit spots engulfing them almost completely
(FIG.~\ref{fig-morph} (a)).
The adjacent T zone is crescent-shaped but tends to form a closed shell around the B cells.
The T zone is basically determined as non-B zone and T cells occupy the remaining space by random migration
around the follicle. However, they do not diffuse freely in the whole space due to the chemoattraction
to the exit spots. Thus this toy model of PLF formation generates a tissue in flow equilibrium with
dynamically generated sources of cell attraction and keeps B and T cells separated.

\subsection{Effect of receptor internalization}
\label{sec-recint}
%
When studying a stable follicle including chemotaxis towards exit spots
the effect of CXCR5 internalization dynamics on the follicle shape
is not visible.
In order to isolate the structural effect of CXCR5 receptor internalization,
the chemoattraction to the exit spots 
is switched off.
Then a flow equilibrium of B cells
is no longer possible as the efflux of cells is strongly suppressed (not shown).
This reflects that randomly migrating cells rarely find the exit spots.
Consequently, cell influx and efflux are completely shut off which imposes
a constant B cell number. This allows to study the FRC-FDC dynamics
together with the internalization dynamics of CXCR5 alone.
It is found that the internalization dynamics on its own 
is a source of instability for the system
as the forming follicle cannot maintain its shape (FIG.~\ref{fig-morph} (b)).

Note, that this result is not affected by considering the internalization
dynamics under B cell flow conditions. This can be investigated by compensating
the low efflux of B cells, which arises due to switched off chemoattraction
to the exit spots, by distribution of the exit spots over a large area.
The instability is even stronger because the outflowing B cells act as an additional sink
for CXCL13 by taking surface bound CXCL13 out of the tissue.

The effect of receptor internalization can be understood by comparing the results
to a simulation without CXCR5 internalization.
If the binding state of the chemoattractant receptors is frozen
the cells would migrate like being in a potential attracting them to FDC. This
is rather independent of the B cell-induced dynamics between FDC and FRC because the migration
is much faster such that the B cell distribution can equilibrate. 
Thus the distribution of B cells and FDC would match each other and the follicle
would be stable. 
The internalization dynamics can however change the 'potential' formed
by the chemoattractant on a time scale comparable to the cell migration
(FIG.~\ref{fig-morph}(c)). As the cells act as sinks
for the chemoattractant they are always generating local gradients 
away from their current position.
This cannot be overcome by diffusion of the chemoattractant which is slower 
than the cell migration on the corresponding cellular length scale. 
Note that the instability only occurs for a sufficiently large number of
B cells because otherwise the sink for the chemoattractant is too small
to change CXCL13 gradients significantly.

The instability of the follicle shape is amplified by the FRC-FDC dynamics. As the B cells
spread out preferentially at the boundary of the follicle, where CXCL13 concentrations are rather low,
they extent beyond the follicle border and increase the B cell density outside the follicle.
Thus, the area covered by the B cells is bigger 
than expected from a dense packing of the B cells in the FDC area.
This culminates in the generation of new FDC.
As on one hand B cells are very motile and on the other
hand FDC need some time to dedifferentiate to FRC,
on long term this results in an FDC network which is bigger
than the volume required by the number of B cells.
Thus the whole PLF becomes unstable in shape
following the changed 'center of mass' of the FDC network
as determined by the concentration peaks of CXCL13.
%
\section{Discussion}
A model architecture was presented that allows to simulate fast cell migration taking into
account subtle effects of chemoattractants and cell influx and efflux from tissues.
Cells are represented as visco-elastic objects. They interact with adjacent cells
by passive and active forces.
The model allows to simulate detailed mechanics of single cells and individual coupling of internal
cell dynamics to cell mechanics as well as to contact- and long-range-interactions.
The use of an underlying regular triangulation
permits a continuous description of cell positions and motility. 
It provides an efficient way of defining the neighborhood
topology for cells of different size in tissue of different density.
The advantage of the regular triangulation
is to provide an effective method to implement fluxes of highly motile cells
by dynamical modifications of the triangulation.
In contrast, lattice-based architectures have difficulties to model fast cell migration.
Either complicated rules are needed or a lattice-gas model
is required \cite{Deutsch:2003}. In both cases the parameters
coupled with the modifications are not easily converted to observable
quantities. Using a physical representation of cells
allows to directly access the involved parameters by
experiment. This consequently requires a lattice-free description
as implemented in the present model based on a regular triangulation.
The triangulation is complemented by a grid for solubles like chemokines.
The reaction-diffusion-equations are solved on that grid. Thus, the model
acquires a hybrid model architecture.
The model is based on
parameters with physiological well-defined meaning such that the number
of parameters is considerably reduced if the corresponding experimental values
are known. This is, indeed, the case for most parameters in the present test-application.

The biomechanics of cellular aggregates of fast migrating cells has three components: passive mechanical
interaction described by JKR forces, active forces exerted on extra-cellular matrix, and active forces
directly acting between cells. 
In general the active forces dominate the JKR forces which allows cell
movement in dense configurations. 
The speed distribution with high friction
and consequently large active forces allows cells to more easily overcome the repulsion generated by JKR
forces. Thus, the high friction regime mimics large deformations of cells
without explicitely taking into account the shape of migrating lymphocytes which elongate in the direction
of motion \cite{Miller:2002,Miller:2003,Wei:2003,Okada:2005,mehe:2005}. Alternatively,
the elastic modulus $E_i$ of cells could be scaled to appropriate values below the physiologic values
and correspondingly decreased adhesion parameters $\sigma_i$.

The constriction ring model takes into account lateral force generation by actively migrating
cells which are ignored in other force-based models that consider only the generated net force
\cite{Palsson:2000,Dallon:2004}. It is shown that the cell interaction due to the exchange
of active forces is the major determinant of the width of the speed distributions.
Thus this interaction might be interpreted as a Brownian motion-like 
fluctuation of the cell speed also it is deterministic.
The parameter coupled with the constriction ring is the pressure
the pressure $p^*_i$ \refeq{eq-ring}. The force
$F^{\rm act}_i\left(\phi_i\right)$ exerted to the extra-cellular
matrix may be either generated by the integrated pressure of the constriction ring model
directly exerted to the matrix
or by the filopodia dynamics of the classical three-step migration model
(reviewed in \cite{Lauffenburger:1996,Mitchison:1996}). Recent data demonstrates
that both migration modes act in lymphocytes \cite{Yoshida:2006} although the contribution of
the constriction ring is greater during chemotactic responses of migrating
lymphocytes \cite{Friedl:2000,Friedl:2001,Yoshida:2006}. Especially, the characteristic time
of the cytoskeleton dynamics of the ring coincides nicely with the persistence time
\cite{Paluch:2005,Yoshida:2006}.
The persistence time, the time interval during which the direction
of active forces is constant, is likely to be an
intrinsic feature of the underlying cytoskeleton dynamics
\cite{Cunningham:1995,Keller:1998,Paluch:2005,Paluch:2006,Yoshida:2006}.

The test-application for the full model involves a simple approach to the formation
of PLF and has proven the functionality of the model.
More specifically, the motility properties of the cells 
could be adjusted to {\it in vivo} motility
data of lymphocytes in lymphoid follicles.
The model generates homeostatic follicles in a flow equilibrium 
of reasonable size of few 
100 \MICRON~~\cite{Bhalla:1981,Ohtani:2003,Halleraker:1994,Kasajima-akatsuka:2006,Belz:1998}
demonstrating that the model architecture is able to describe systems that are in
a cell flow equilibrium of fast migrating cells.

The effect of CXCR5 receptor internalization was demonstrated under non-flow conditions.
Within the investigated parameter range (see appendix)
the dynamics of CXCR5 and CXCL13 leads to a modification
of the concentration profile such that cells temporarily migrate away from sources. This leads to
a frequent change in shape of the cell aggregate. Such a behavior was not observed 
{\em in vivo} \cite{Wei:2003}. Therefore either internalization of CXCR5 does not occur
in the PLF system or mechanisms exist that counter-balance the shape dynamics.
Higher CXCL13 concentrations can overcome this effect as the internalization cannot act
as a sufficient sink to locally invert the gradient of the chemoattractant (not shown).
However, the cells then have strongly downregulated receptors such that they
loose there chemotactic sensitivity, i.e.~cells start to migrate randomly
preventing efficient aggregation on long terms.
We observed a stable follicle shape under flow conditions
when the B cells chemotax to the exit spots without internalization dynamics. 
Thus,
chemotaxis to the exit
spots works as an independent attractor for B cells that can compenstate for
locally inverted gradients of CXCL13 resulting from the receptor internalization
of CXCR5.

The main purpose of this article was to establish a simulation platform suitable
for modeling of homeostatic tissue dynamics on a cellular and subcellular level
involving large numbers of fast migrating cells.
The results have demonstrated that the proposed model design can 
cope with the complexity that occurs in tissue exhibiting
a flow equilibrium of fast migrating cells. In particular, it clearly
separates the various time and length scales and allows to localize
the origin of emerging properties on the tissue level as well as on the
cellular and molecular level. We, therefore, 
consider this simulation tool to be a suitable instrument for
the analysis of morphogenesis of highly dynamic tissue.
However, concerning the
application to follicle formation, the assumed mechanisms are only subparts
of full PLF formation dynamics that need to be
refined in order to reproduce microanatomical data and 
to make realistic and quantitative predictions, which
is left for future research.

\section*{Acknowledgments}
FIAS is supported by the ALTANA AG. Meyer-Hermann is supported by
the EU-NEST project MAMOCELL within FP6.
\appendix*

\section{Estimate parameters for chemoattractant reaction-diffusion system}
The parameters used in the ODE system \refeq{eq-desens} are not known but can be estimated from data of similar systems.
The dissociation constant $K_{\rm d}$ for chemoattractants and their receptors are
mostly measured for other chemoattractants than the ones
used here (CCL21 and CXCL13) \reftable{tab-param}.
The values for $K_{\rm d}$ range from 0.2 nM to 
5 nM \cite{Yoshida:1998,Lin:2004,Pelletier:2000,Slimani:2003}.
The dissociation constant for CCR7 with CCL21 has been measured to be 1.6 nM \cite{Willimann:1998}.
Considering the range of these values it is likely that $K_{\rm d}$ 
adopts values in this range in the present system.
A less favorable situation exists for the reaction rate constant
$k_{\rm on}$ ($k_{\rm off} = k_{\rm on}K_{\rm d}$).
Only few data exists which spread over several orders of magnitude. 
Values for the association rate $k_{\rm on}$ are available for CXCL12 binding to fibronectin
\cite{Pelletier:2000} ($2.5\cdot10^5 \,{\rm M^{-1}s^{-1}}$) .
The off rate $k_{\rm off}$ measured for CXCL12 binding to fibronectin 
is $6.5\cdot10^{-3} \,{\rm s^{-1}}$.

The two rate constants $k_{\rm i}$ and $k_{\rm r}$
associated with receptor internalization and recycling (see \refeq{eq-desens}) can be estimated 
from experimental data on receptor desensitization and
resensitization experiments which are reviewed recently \cite{Neel:2005}. 
It is assumed that all non-internalized receptors have bound the ligand
when high chemoattractant concentrations far above $K_{\rm d}$
are used as done in the experiment.
The only equation that is left for the desensitization process 
is then \refeq{eq-desens} ($R=0$) 
\be  
  \dot R^*       &=& k_{\rm i}(R_{\rm tot} - R^*) - k_{\rm r}R^*
\ee
with the solution
\be\label{eq-kes}
  R^*(t) &=& \frac{R_{\rm tot}}{1+k_{\rm r}/k_{\rm i}}\left\{1 - \exp{[-(k_{\rm i}+k_{\rm r})t]} \right\}.
\ee
The internalized receptor fraction $r^*$ at equilibrium becomes
\be
  r^* = \frac{R^*(t\rightarrow\infty)}{R_{\rm tot}} &=& \frac{1}{1+k_{\rm r}/k_{\rm i}}.
\ee
For the resensitization process we set $k_{\rm i}=0$ and start from 
$R^*(t=0)=r^*R_{\rm tot}$. In the absence of the ligand the dynamics
for the internalized receptor during resensitization becomes
\be
  R^*(t) &=& r^*R_{\rm tot}\exp{[-k_{\rm r}t]}.
\ee
With typical values of $r^* = 0.3\ldots 0.8$ for desensitization
and typical recycling times of 60--180 minutes (to get $r^* \sim 0.2$ upon resensitization
\cite{Neel:2005}) one arrives at
\begin{eqnarray*}
  k_{\rm r} &=& 1 \cdot 10^{-4} \,\ldots 7\cdot10^{-3}\, {\rm s^{-1}}\\
  k_{\rm i} &=& 5 \cdot 10^{-5} \,\ldots 3\cdot10^{-2}\, {\rm s^{-1}}.
\end{eqnarray*}
Assuming that the internalization process is not in steady state -- and solving \refeq{eq-kes}
numerically -- doesn't change the results very much compared to the experimental uncertainty.
The numerical results of PLF-formation
are not sensitive to these parameters (data not shown).

$R_{\rm tot}$ is not known explicitly.
From similar receptors the number of CCR7 molecules
on T cells has been estimated to be $10^5$ per cell \cite{Willimann:1998}
and $10^4$ for B cells as indicated by the studies that find
a factor 10 difference of CCR7 levels between B and T cells \cite{Okada:2002}.
Note that the values cited above have to be converted into the receptor concentration
$R_{\rm tot}$ based on the cell densities of the corresponding lymphocyte type.

\begin{table}
 \centering
 {
 \begin{tabular}{lllll} \hline
   $\kappa$     & & $2 \cdot 10^{-4}\,{\rm s^{-1}}$ & & \cite{Bar-even:2006}\\
   $k_{\rm i}$  & & $5 \cdot 10^{-5} \ldots 3\cdot10^{-2} \,{\rm s^{-1}}$ & &\cite{Neel:2005} \\
   $k_{\rm r}$  & & $1 \cdot 10^{-4} \ldots 7\cdot10^{-3} \,{\rm s^{-1}}$ & &\cite{Neel:2005} \\
   $K_{\rm d}$  & & $0.2\ldots 5\,{\rm nM}$                               & &\cite{Willimann:1998,Yoshida:1998,Lin:2004,Pelletier:2000,Slimani:2003} \\
   $k_{\rm on}$ & & $2.5\cdot10^5 \ldots 10^8 \,{\rm M^{-1}s^{-1}} $      & &\cite{Pelletier:2000} \\
   $k_{\rm off}$ & & $10^{-4} \ldots 1 \,{\rm s^{-1}}  $                  & &(from $K_{\rm d}$ and $k_{\rm on})$ \\
   $Q$          & & $2.5\cdot 10^1\ldots 10^4\,{\rm s^{-1}}$              & &\cite{Vissers:2001,Hu:2004} \\ 
   $R_{\rm tot}$                 & & $10^4$--$10^5$ & & \cite{Willimann:1998,Okada:2002} \\
\hline
 \end{tabular}
 }
 \vspace{0.3cm}
 \caption[Receptor internalization parameters]{\label{tab-param}
    Parameter for the equation system \refeq{eq-desens}.}
\end{table}


\end{document}